\documentclass{jfm}
\usepackage{amsmath,color,graphicx,amssymb}
\usepackage{bm}
\usepackage{natbib}
\usepackage{amsbsy}
\usepackage{latexsym}
\usepackage{subfigure}
\usepackage[normalem]{ulem}
\usepackage[english]{babel}
\usepackage{wasysym}
\usepackage{epsfig} 
\usepackage{epsf,psfig}
\usepackage{epstopdf}
\usepackage{multirow}
\usepackage{array}
\usepackage{float}
\usepackage[english]{babel}
\usepackage{verbatim}
\usepackage{soul}
\usepackage[toc,page]{appendix}
\usepackage{setspace}
\usepackage{mathtools}

\definecolor{gray}{rgb}{0.5,0.5,0.5}
\definecolor{black}{rgb}{0,0,0}
\definecolor{purple}{rgb}{0.57,0,0.86}
\definecolor{greenish}{rgb}{0.2,0.7,0.2}
\definecolor{orange}{rgb}{1,0.5,0}

\newcommand\bnabla{\boldsymbol{\nabla}}

\newcommand\mb{\mathbf}
\newcommand\bsy{\boldsymbol}
\newcommand\mc{\mathcal}

\def\beq{ \begin{equation}}
\def\eeq{\end{equation}}
\def\beqar{ \begin{eqnarray} }
\def\eeqar{ \end{eqnarray} }

\def\vK{von K\'{a}rm\'{a}n }

\begin{document}


\title{Unsteady Propulsion by an Intermittent Swimming Gait}

\author[E. Akoz \& K.W. Moored]
{By E\ls M\ls R\ls E\ns A\ls K\ls O\ls Z$^1$ \ns
\and K\ls E\ls I\ls T\ls H\ns W.\ns  M\ls O\ls O\ls R\ls E\ls D,$^1$}

\affiliation{$^1$Department of Mechanical Engineering and Mechanics\\
Lehigh University,
Bethlehem, PA 18015, USA\\
[\affilskip]
}


\maketitle


\begin{abstract}

Inviscid computational results are presented on a self-propelled swimmer modeled as a virtual body combined with a two-dimensional hydrofoil pitching intermittently about its leading edge. \cite{Lighthill1971} originally proposed that this burst-and-coast behavior can save fish energy during swimming by taking advantage of the viscous Bone-Lighthill boundary layer thinning mechanism.  Here, an additional inviscid Garrick mechanism is discovered that allows swimmers to control the ratio of their added mass thrust-producing forces to their circulatory drag-inducing forces by decreasing their duty cycle, $DC$, of locomotion.  This mechanism can save intermittent swimmers as much as 60\% of the energy it takes to swim continuously at the same speed.  The inviscid energy savings are shown to increase with increasing amplitude of motion, increase with decreasing Lighthill number, $Li$, and switch to an energetic cost above continuous swimming for sufficiently low $DC$.  Intermittent swimmers are observed to shed four vortices per cycle that form into groups that are self-similar with the $DC$.   In addition, previous thrust and power scaling laws of continuous self-propelled swimming are further generalized to include intermittent swimming.  The key is that by averaging the thrust and power coefficients over only the bursting period then the intermittent problem can be transformed into a continuous one.  Furthermore, the intermittent thrust and power scaling relations are extended to predict the mean speed and cost of transport of swimmers.  By tuning a few coefficients with a handful of simulations these self-propelled relations can become predictive.  In the current study, the mean speed and cost of transport are predicted to within 3\% and 18\% of their full-scale values by using these relations.
\end{abstract}

%
%
%


\section{Introduction}
Many fish such as cod \cite[]{Videler1981}, saithe \cite[]{Videler1982}, zebra danios \cite[]{Muller2000}, cormorant \cite[]{Ribak2005} and koi carps \cite[]{Wu2007} swim by using a combination of an active swimming phase and a passive coasting phase known as burst-and-coast or burst-and-glide swimming.  This intermittent swimming gait adopted by fish, has been predicted by dynamical models to yield an energy savings to swim a given distance of over 50\% \cite[]{Weih1974,Weihs1980}.   These predictions are based on the Bone-Lighthill boundary layer thinning hypothesis, which proposes that there is an increase in skin friction drag on a fish body when it undulates \cite[]{Lighthill1971}.  Under these conditions, a fish may then reduce its overall skin friction drag by interspersing an undulation phase with a coasting phase.  The skin friction increase is caused by the thinning of the boundary layer from the body normal velocity component that is present during undulation.  In fact, Lighthill originally estimated that the skin friction drag on a swimming body can be up to a factor of 5 times larger than a gliding body \cite[]{Lighthill1971}.  Others estimated that the factor could be in a range of 4-10  \cite[]{Webb1975,Videler1981,Wu2007}.  More recently, detailed analysis confirms the boundary layer thinning hypothesis, but estimates the skin friction drag increase to be much more modest than originally proposed and in the range of a 20--90\% drag increase \cite[]{Ehrenstein2013,Ehrenstein2014}.  
 
Given this range of skin friction increase, the viscous mechanism proposed by Bone and Lighthill (\citeyear[]{Lighthill1971}) is not sufficient to create a 50\% savings in energy by itself.  Yet, a 45\% energy savings was indirectly estimated from DPIV experiments that measured the thrust impulse imparted to the wake of koi carps \cite[]{Wu2007}.  Also, a 56\% energy savings for an intermittent fish-like swimmer was directly calculated by using a two-dimensional Navier-Stokes solver \cite[]{Chung2009}.

Motivated by these observations, we hypothesize that there is an inviscid mechanism that can account for the majority of the energetic benefit seen in intermittent swimming for parameters typical of biology.  To examine this hypothesis, we show that by using inviscid computations there is an energy savings for intermittent swimming as compared to continuous swimming at the same swimming speed.  Then we consider three questions: How do the forces and energetics of intermittent swimming scale with the swimming parameters, what is the nature of the invicsid mechanism that leads to the observed energy savings, and what are the limitations to the benefit?

\section{Approach and Numerical Methods}
\subsection{Problem Formulation and Input Parameters}
\begin{figure}
\centering
\includegraphics[width=0.99\textwidth]{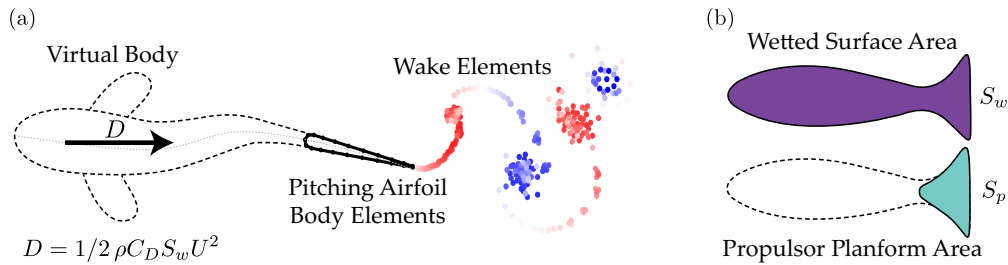}
\caption{ (a) Schematic of the idealized self-propelled swimmer. Dashed line represents the virtual body. There is a drag force, $D$, acting on the hydrofoil that represents the effect of the virtual body. The wake element end points are colored red and blue for counter clockwise and clockwise circulation, respectively. (b) Schematic representation of the wetted surface area, $S_{w}$, (purple shaded area) and the propulsor area, $S_{p}$, (teal shaded area) for a generic swimmer.}
\label{schematic}
\end{figure}

\begin{figure}
\centering
\includegraphics[width=0.95\textwidth]{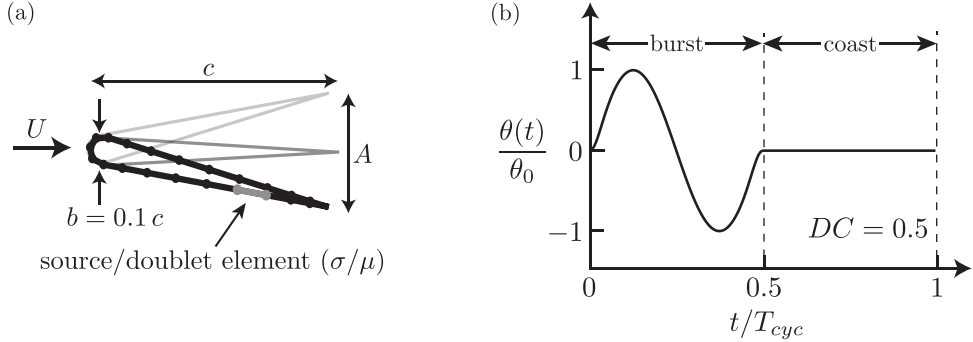}
\caption{(a) Geometric and numerical parameters for the teardrop hydrofoil. (b) Normalized pitching angle as a function of normalized time for an intermittent swimmer with $DC=0.5$. }
\label{signal}
\end{figure}

A computational study is performed to compare the performance and flow fields of idealized continuous and intermittent swimmers.  A self-propelled swimmer is used to model the problem that is a combination of a virtual body and a two-dimensional hydrofoil pitching about its leading edge (Figure \ref{schematic}a).  The virtual body is not present in the computational domain, however, its presence is modeled as a drag force, $D$, that acts to resist the motion of the swimmer.  The magnitude of the drag force is determined from a drag law based on high Reynolds number swimming conditions \cite[]{Munson1990},
\begin{align}\label{eq:draglaw}
D=1/2\, \rho C_D S_w U^2
\end{align}

\noindent where $\rho$ is the density of the fluid, $C_D$ is the coefficient of drag of the virtual body, $S_w$ is the combined wetted surface area of the virtual body and the propulsor (Figure \ref{schematic}b), and $U$ is the speed of the swimmer.  The planform area of the hydrofoil, $S_p$, is defined as the chord length, $c$, multiplied by a unit span length.  The chord length is fixed to $c = 0.07$ m, which is typical of species that use intermittent swimming gaits \cite[]{Beamish1966, Videler1981}.  An area ratio can then be formed as the ratio of the wetted surface area to the propulsor planform area, 
\[S_{wp} \equiv \frac{S_w}{S_p}.\]
The area ratio, used to calculate $S_w$ in the drag law, is selected to be $S_{wp} = 8$, which is also typical for intermittently swimming species \cite[]{Beamish1966, Videler1981, Videler1991, Webber2001}. In addition, the pitching hydrofoil has a teardrop cross-sectional shape \cite[]{Marais2012} with a thickness-to-chord ratio of $b/c = 0.1$ (see Figure \ref{signal}a).

Previous analytical and computational models of intermittent swimming assume a higher $C_D$ during a bursting phase than a gliding phase by referring to the Bone-Lighthill boundary layer thinning hypothesis \cite[]{Weih1974,Muller2000,Wu2007,Chung2009}. However, in this study we set $C_D$ to be a \textit{constant} value for both the bursting and gliding phases of intermittent swimming.  This allows us to probe the hypothesis that the energy savings observed in intermittent swimming is not wholly due to the boundary layer thinning hypothesis. In other words, when $C_D$ is fixed regardless of whether a swimmer is oscillating its fin or gliding, then any observed energy savings comes purely from potential flow mechanisms and is not associated with a skin friction rise due to the oscillating fin and/or body. In this study, three different \textit{constant} drag coefficients, $C_D = 0.005,\, 0.01,$ and $0.05$, are used to represent the range of average drag coefficients observed in biology \cite[]{Videler1981, Videler1982, Webb1984, Anderson2001, Wu2007, Ehrenstein2013, Ehrenstein2014}.

The drag coefficient and the area ratio can be grouped together into a combination parameter known as the Lighthill number \cite[]{Eloy2013}, 
\begin{align}
Li = C_D\, S_{wp}
\end{align}

\noindent The Lighthill number characterizes how the body and propulsor geometry affects the balance of thrust and drag forces on a swimmer. For example, for a fixed thrust force and propulsor area during self-propelled continuous swimming, the swimming speed is inversely proportional to the Lighthill number \cite[]{Moored2017}.  If $Li$ is high then a swimmer produces high drag leading to low self-propelled swimming speeds and \textit{vice versa}.  In the present study, the Lighthill numbers examined are $Li = 0.04,\, 0.08,$ and $0.4$, which fall in a range typical of biology \cite[]{Eloy2013}.

The virtual body also must be given a mass, $m$, which can be non-dimensionalized with the added mass of the hydrofoil propulsor, that is,
\begin{align}
m^* \equiv \frac{m}{\rho S_p c}.
\end{align}

\noindent The non-dimensional mass, $m^*$, affects not only the acceleration of a swimmer due to any net time-averaged forces over a period, but it also affects the magnitude of the surging oscillations that occur within a period due to the unsteady forcing of the pitching hydrofoil. By using the mass of a typical intermittent swimmer and the dimensions of the pitching hydrofoil, the non-dimensional mass is set to $m^* = 3.86$ throughout this study.  Even though this mass ratio is based on the biology, previous self-propelled numerical solutions show that the time-averaged forces and energetics are effectively independent of the amplitude of surging oscillations for $m^* > 1$ \cite[]{Moored2017} even though the instantaneous forces can be affected \cite[]{Wen2013}.

The kinematic motion is parameterized with a pitching frequency, $f$, and a peak-to-peak trailing-edge amplitude, $A$, reported as a non-dimensional amplitude-to-chord ratio, $A^* = A/c$, and a duty cycle, $DC$.  The amplitude-to-chord ratio is related to the maximum pitch angle, that is, $\theta_0 = \mbox{sin}^{-1}\left(A^*/2\right)$.  The degree of intermittency of swimming is captured in the duty cycle, which is,
\begin{align}
DC=\frac{\text{burst period}}{\text{total cycle period}}=\frac{T_{\text{burst}}}{T_{\text{cycle}}}.
\end{align}
\begin{table}
   \begin{center}
	\begin{tabular}{lccccccccccc} 
  	 Continuous Swimmers &  & & &  & &  & & & & & \vspace{-5pt} \\  \hline 
	 & & & & &  & & & & & & \vspace{-10pt} \\
         $Li$ & & 0.04 & 0.08 & 0.4 &  & &  & & & &    \\
	$f$  (Hz)& & 0.2 & 0.4 & 0.6 & 0.8 & 1  &&  &  &  & \\
	$DC$& & 1 & &  & &  & &  &  &  & \\
	$A^*$& & 0.3 & 0.4 & 0.5 & 0.6  & 0.7 &   & & & &   \\
	$\theta_0$  (deg.)& & 8.6  & 11.5  & 14.5 &  17.5 &  20.5 & & && & \\ \\
	Intermittent Swimmers &  & & &  & &  & & & & &  \vspace{-5pt} \\  \hline 
	 & & & & &  & & & & & & \vspace{-10pt} \\
         $Li$ & & 0.04 & 0.08 & 0.4 &  & &  & & & &    \\
	$f$  (Hz)& & 1 &  &  &  &  &  &  &  & & \\
	$DC$ & & 0.2 & 0.3 & 0.4 & 0.5 & 0.6  & 0.7 & 0.8 & 0.9 & & \\
	$A^*$& & 0.3 & 0.4 & 0.5 & 0.6  & 0.7 &   & & & &   \\
	$\theta_0$  (deg.)& & 8.6  & 11.5  & 14.5 &  17.5 &  20.5 &  & & & & \\ \\
	Fixed Parameters&  & & &  & &  & & & & &   \vspace{-5pt} \\  \hline 
	 & & & & &  & & & & &  & \vspace{-10pt} \\
         $m^* = 3.86$ & $S_{wp} = 8 $ &   &  & &   & &  & & & & \\ \\
	\end{tabular} 
   \end{center}
  	\caption{Simulations parameters used in the present study.}
 	\label{tab:parameters}
\end{table}

\noindent Figure \ref{signal}b shows a characteristic pitching motion with the burst, coast and total cycle periods defined.  The intermittent motion is a combination of a sinusoidal pitching motion for the burst period and it is followed by a fixed pitch angle of $\theta = 0$ for the duration of the coast period.  The total cycle period is simply the addition of the burst and coast periods.  The combined burst and coast pitching motions about the leading edge of the hydrofoil is then defined as, 
\begin{align} \label{eq:signal}
\theta(t) &= 
\begin{dcases}
    y_{s}(t) \left[\theta_0 \, \mbox{sin} \left( 2 \pi f t\right) \right], &  0 \leq t \leq T_{\text{burst}}\\
    0,              &  T_{\text{burst}} \leq t \leq T_{\text{cycle}}
\end{dcases} \\
\mbox{where} \quad y_s(t) &= 
\begin{dcases}
    -\mbox{tanh}(m\, t)\, \mbox{tanh}\left[m\, (t-1)\right], &  DC < 1\\
    1, & DC = 1\\
\end{dcases} \label{eq:smoothing}
\end{align}

\noindent Equation (\ref{eq:signal}) defines a reference signal where $0 \leq t \leq T_{\text{cycle}}$. The signal used in the simulations has $N_{cyc}$ repetitions of this reference signal.  Here, $T_{\text{burst}} = 1/f$ and $T_{\text{cycle}} = T_{\text{burst}}/DC$.  

 In order to obtain discretization independent solutions as the time step size is reduced, the discontinuous angular rates and accelerations at the junction of the burst phase and coast phase must be smoothed.  To do this, a hyperbolic tangent envelope function, $y_s(t)$, is multiplied with the sinusoidal burst signal and is defined in (\ref{eq:smoothing}). This function modifies the slope of the sine wave at $t/T_{\text{burst}}=0$ and $t/T_{\text{burst}}=1$ to ensure a desingularized smooth junction with the coast phase where $m$ controls the radius of curvature of the junction. In the current study, $m = 9$ is used. Additionally, if $DC = 1$, then the signal (\ref{eq:signal}) reverts to a continuous sinusoidal signal. In the current study the duty cycle ranges from $DC = 0.2$ to $DC = 1$ in $0.1$ increments.  A summary of the input parameters used in the current study are in Table \ref{tab:parameters}.


\subsection{Numerical Methods}
To model a high Reynolds number fluid flow, the field around the hydrofoils is assumed inviscid, irrotational (except on the boundary elements) and incompressible. For an incompressible and irrotational fluid the continuity equation reduces to Laplace'�s equation, $\nabla^2{\phi^*}=0$, where $\phi^*$ is the perturbation potential in an inertial frame fixed to the undisturbed fluid. In addition, the no-flux boundary condition must be satisfied on the body's surface, $\mc{S}_b$ that is, 
\begin{align} \label{eq:noflux}
\bnabla \phi^* \cdot \mathbf{n}=0 \quad \quad \text{on } \mc{S}_b
\end{align}
where $\mathbf{n}$ is a vector normal to the body's surface, and the velocity is $\mb{u} = \bnabla{\phi^*}$. Additionally, the velocity disturbances created by the motion should satisfy the far-field boundary condition and decay far from the body,
\begin{align}\label{eq:farfield}
\lim_{r\to\infty} (\bnabla{\phi^*})=0
\end{align}
where $\mathbf{r}= x\, \hat{\mb{x}} + z\, \hat{\mb{z}} $ measured from a body-fixed frame.

Following \cite{Katz1991} and \cite[]{Quinn2014}, the general solution for the perturbation potential, also known as the boundary integral equation is explicitly stated as a combination of a distribution of sources of strength $\sigma$ and doublets of strength $\mu$ on the body surface $\mc{S}_b$ and a distribution of doublets of strength $\mu_w$ on the wake surface $\mc{S}_w$, 
\begin{align} \label{IntPertPot}
\phi_i^*(\mb{r}) & = \oiint_{\mc{S}_b} \left[ \sigma(\mb{r_0}) \; G(\mb{r};\mb{r_0}) - \mu(\mb{r_0}) \; \mb{\hat{n}} \cdot \bsy{\nabla} G(\mb{r};\mb{r_0})\right] \; d\mc{S}_0 -\oiint_{\mc{S}_w} \mu_w(\mb{r_0}) \; \mb{\hat{n}} \cdot \bsy{\nabla} G(\mb{r};\mb{r_0}) \; d\mc{S}_0  \\ 
& \mbox{where,} \nonumber \\
\sigma & = \mb{\hat{n}} \cdot \bsy{\nabla} \left(\phi^* -  \phi_i^*\right) \label{Source} \\
-\mu & = \phi^* - \phi_i^* \\
-\mu_w & = \phi_+^* - \phi_-^*
\end{align}

\noindent Here $\phi^*_i$ is the internal potential, $\mb{r_0}$ is the location of a source or doublet, $\mb{r}$ is the target point, and the Green's function for the two-dimensional Laplace equation is $G(\mb{r};\,\mb{r}_0) = \frac{1}{2 \pi} \text{ln} |\mb{r} - \mb{r_0}|$. Conveniently, doublets and sources both implicitly satisfy the far-field boundary condition (\ref{eq:farfield}). The problem is then reduced to determining the source distribution, $\sigma$, and the doublet distribution, $\mu$, over the surfaces such that the no-flux boundary condition (\ref{eq:noflux}) is satisfied. Here a Dirichlet formulation is used to enforce the boundary condition by fixing the potential within the body surface to be equal to zero, that is, $\phi^*_i = 0$.  This leads to the potential field at the surface of the body equated to the local doublet strength as $-\mu = \phi^*_b$ and the local velocity normal to the body surface equated to the source strength as $\sigma = \mb{\hat{n}} \cdot \bsy{\nabla} \phi^*_b= \mb{\hat{n}} \cdot (\mb{u_{rel}} + \mb{U})$. The velocity of the center of each source element is $\mb{u_{rel}}$, which is relative to a body-fixed frame of reference located at the leading edge of the hydrofoil and $\mb{U}$ is the velocity of the body-fixed frame with respect to the undisturbed fluid. 

To solve the problem numerically, surface $\mc{S}_b$ is discretized into $N$ source and doublet boundary elements and surface $\mc{S}_w$ is discretized into $N_w$ doublet boundary elements.  Each line element is formed of a constant strength distribution of point sources or doublets.  In addition, the no-flux boundary condition (equation (\ref{IntPertPot}) with $\phi^*_i = 0$) is enforced at $N$ discrete collocation points that are located along the normal vector of each body boundary element and inside the body by a distance of 15\% of the body half-thickness measured at the center of each element.  Also, an explicit Kutta condition is applied by introducing a wake panel at the trailing edge with a strength $\mu_{w,TE} = \mu_{t,TE} - \mu_{b,TE}$ where $\mu_{t,TE}$ and $\mu_{b,TE}$ are the strengths of the top and bottom body elements at the trailing edge, respectively.  The trailing-edge wake element is oriented along a line that bisects the upper and lower body surfaces at the trailing edge.  The length of the element is set to $0.4\, U \Delta t$ where $\Delta t$ is the time step for the computations.  


To satisfy the Kelvin circulation theorem a wake shedding procedure must be applied.  At every time step, the first wake panel is `shed' by advecting the element downstream by a distance of $U \Delta t$ while its strength remains fixed for all time at $\mu_{w,TE}$ from the previous time step.  Subsequently, a new trailing edge wake element is formed and the shed wake panel is further advected with the local induced velocity field from the other wake and body elements.  During this rollup process, the endpoints of the doublet elements, which are mathematically equivalent to point vortices, must be desingularized for the numerical stability of the solution.  Following Krasny (\citeyear[]{Krasny1986}) the induced velocity on a wake element from other doublet elements is then calculated with a desingularized Biot-Savart law,
\begin{align}
\mb{u}(\mb{r}) &= \frac{\Gamma}{2 \pi} \frac{\hat{\mb{y}} \times \left( \mb{r} - \mb{r_i}\right) }{\left| \mb{r} - \mb{r_i}\right|^2 + \delta^2} \\ &\text{and} \nonumber \\
\mb{r} - \mb{r_i} &= (x- x_i)\, \hat{\mb{x}} + (z-z_i)\, \hat{\mb{z}}  \nonumber 
\end{align}

\noindent Here $x_i$ and $z_i$ are the positions of the $i^{th}$ endpoint of the doublet elements.  The desingularized point vortex circulation is related to its doublet element strength as $\Gamma = -\mu$.  As the desingularization parameter, $\delta$, approaches zero then the classical Biot-Savart law is recovered. In the current study, $\delta/c = 6 \times 10^{-2}$ is selected for the desingularization.

A matrix form of the boundary integral equation (\ref{IntPertPot}) is constructed with $\phi^*_i = 0$, the explicit Kutta condition applied, and the wake element representation described above governing the development of the free shear layer.  The linear equations are solved to determine the unknown body doublet element strengths and subsequently the perturbation potential field $\phi^*(\mb{r})$ and the velocity field $\mb{u}(\mb{r})$.  Then the perturbation velocity on the surface of the body is found by a local differentiation of the perturbation potential,
\begin{align}
\mb{u_b} = \bnabla \phi_b^* = \frac{\partial \phi^*}{\partial s} \; \mb{\hat{s}} + \frac{\partial \phi^*}{\partial n} \; \mb{\hat{n}} = - \frac{\partial \mu}{\partial s} \; \mb{\hat{s}}  +  \sigma \; \mb{\hat{n}},
\end{align}

\noindent where $\mb{\hat{s}}$ is the tangential vector along the surface of the body.  In addition, the pressure field acting on the body is calculated by using the unsteady Bernoulli equation,
\begin{align} \label{BernBody}
 P_b(x,z,t) = \rho \frac{\partial \mu}{\partial t}\bigg|_{body} + \rho \left(\mb{u_{rel} + \mb{U}} \right) \cdot \mb{u_b} - \rho \frac{\mb{u_b} ^2}{2}.
\end{align}

Finally, the forces acting on the pitching hydrofoil are calculated from a combination of the pressure forces acting on the body and the drag law (\ref{eq:draglaw}),
\begin{equation} \label{Forces}
\mb{F}\left(t \right) =  \int_{\mc{S}_b } -P_b\; \mb{\hat{n}}\; d\mc{S}_b + D\, \hat{\mb{x}}
\end{equation}


To examine intermittent motions the computations are required to model self-propelled swimming conditions.  Consequently, a simple single degree of freedom equation of motion is loosely-coupled to the boundary element fluid solver.   Following Borazjani et al. (\citeyear[]{Borazjani2009}), the velocity of the swimmer at the $(n+1)^{th}$ time step is calculated through forward differencing and the position is calculated by using the trapezoidal rule, 

\begin{align}
U^{n+1} = U^{n} +\frac{F_{x,net}^n}{m} \Delta t
\end{align}

\begin{align}
x_{LE}^{n+1} = x_{LE}^{n} +\frac{1}{2}(U^{n+1} + U^n) \Delta t
\end{align}

\noindent where $F_{x,net}^n$ is the net force acting on the hydrofoil in the streamwise direction at the $n^{th}$ timestep, $x_{LE}$ is the leading edge position of the hydrofoil.  The two-dimensional formulation in the current study has a three-dimensional counterpart that has been extensively validated and used to examine bio-inspired \cite[]{Moored2017a} and biological self-propelled swimming \cite[]{Fish2016}.

\subsection{Output Parameters and Performance Metrics}

There are several output parameters used throughout this study.  For many of them, we examine their mean values time-averaged over an oscillation cycle, which are denoted with an overbar such as $\overline{(\boldsymbol{\cdot})}$.  Mean quantities are only taken after a swimmer has reached the steady-state of its cycle-averaged swimming speed. When this occurs, the steady-state cycle-averaged swimming speed will be described as the mean swimming speed and denoted as $\overline{U}$.  The computations are considered to be at a steady state condition when there is negligible mean net thrust acting on the swimmer, defined as,
\begin{align}
C_{T,net} = \frac{\overline{T}-\overline{D}}{\rho S_p \overline{U}^2} <10^{-4},
\end{align}

\noindent where $\overline{T}$ is the mean thrust force calculated as the streamwise force from the pressure integration term in (\ref{Forces}) alone.  The Strouhal number and the reduced frequency are typical nondimensional frequencies used in bio-propulsion studies and are defined as,
\begin{align}
St=\frac{fA}{\overline{U}} \qquad \qquad \qquad \qquad k= \frac {fc}{\overline{U}}
\end{align}

\noindent The Strouhal number represents the ratio of the cross-stream spacing to the streamwise spacing of the vortices in the wake of a swimmer.  The reduced frequency is a ratio of the time that is takes a fluid particle to traverse the chord of the hydrofoil compared to the period of motion and is a measure of the unsteadiness of the flow.  Since in self-propelled swimming the mean speed is unknown \textit{a priori} then both the Strouhal number and the reduced frequency are also unknown \textit{a priori}.  The time-averaged thrust and power coefficients depend upon the Strouhal number and the reduced frequency and are, 
\begin{align} \label{coeffs}
C_T \equiv \frac{\overline{T}}{ \rho S_p f^2 A^2} && C_P \equiv \frac{\overline{P}}{ \rho S_p f^2 A^2 \overline{U}}
\end{align}

\noindent These coefficients are nondimensionalized with the added mass forces and added mass power from linear theory \cite[]{Garrick1936}.  Additionally, the mean power input to the fluid is calculated as the negative inner product of the force vector and velocity vector of each boundary element.  The ratio of these coefficients leads to the propulsive efficiency, $\eta$, which is linked to the cost of transport, $CoT$,
\begin{align} \label{eff_cot}
 \eta=\frac{\overline{TU}}{\overline{P}}  \qquad \qquad \qquad \qquad CoT=\frac{\overline{P}}{m\overline{U}}
\end{align}

\noindent The propulsive efficiency, $\eta$, is defined as the ratio of useful mean power output to the mean power input to the fluid.  In self-propelled swimming we define this quantity in the potential flow sense \cite[]{Lighthill1971}, that is, the mean thrust force is calculated as the integration of the pressure forces alone.  In this sense the propulsive efficiency is not ill-defined for self-propelled swimming as is the case when the net thrust force is used.  The cost of transport is a measure of the amount of energy it takes to swim a unit distance reported on a per unit mass basis.  It is widely used throughout biological literature \cite[]{Videler1991} and is a useful engineering metric since its inverse is the proportionality constant between the range of a vehicle, $\mathcal{R}$, and its energy density ($\mathcal{E} \equiv \mbox{energy per unit mass} = E/m$), that is, $\mathcal{R} = \left(1/CoT\right) \mathcal{E}$.

To compare the energetic performance of an intermittent swimmer with a continuous swimmer at the same mean speed the normalized cost of transport and efficiency are introduced, 
\begin{align} \label{Norm_cot_eff}
\hat{CoT} \equiv \frac{CoT_{int}}{CoT_{cont}}\bigg|_{\overline{U}} \qquad \qquad \qquad \qquad  \hat{\eta} \equiv \frac{\eta_{int}}{\eta_{cont}}\bigg|_{\overline{U}}
\end{align}

\noindent Here the cost of transport of an intermittent and continuous swimmer are $CoT_{int}$ and $CoT_{cont}$, respectively, and the efficiency of an intermittent and continuous swimmer are $\eta_{int}$ and $\eta_{cont}$, respectively.

\subsection{Discretization Independence}
\begin{figure}
\centering  
\includegraphics[width=0.95\textwidth]{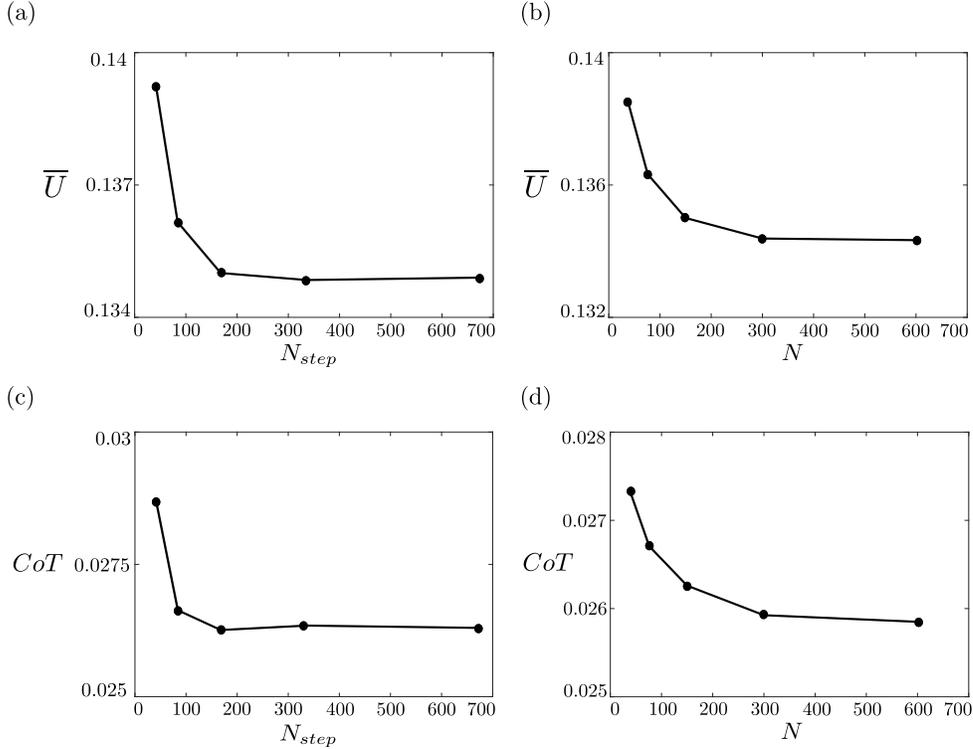}
\caption{Mean swimming speed as a function of the (a) number of time steps during the bursting period and (b) the number of body elements.  Cost of transport as a function of the (c) number of time steps during the bursting period and  (d) the number of body elements.  For these discretization independence simulations the parameters used are $Li = 0.08,\, A^* = 0.5,\, DC = 0.5,$ and $f =1$ Hz.}\label{fig:DiscInd}
\end{figure}

Figure \ref{fig:DiscInd} shows the mean swimming speed and cost of transport as a function of the number of body elements and the number of time steps during the bursting period used in the simulations.  The parameters used are $Li = 0.08,\, A^* = 0.5,\, DC = 0.5,$ and $f =1$ Hz.  It is evident that the mean speed and cost of transport are converging to discretization independent solutions as the number of body elements and time steps is increased.  Both performance metrics change by less than 2\% when $N = 150$ is doubled or $N_{\text{step}} = 150$ is doubled.  Therefore throughout this study these values for the number of body elements and number of time steps during the bursting period will be used.  



\section{Results}
\subsection{Performance} \label{sec:perf}
\begin{figure}
\centering
\includegraphics[width=0.99\textwidth]{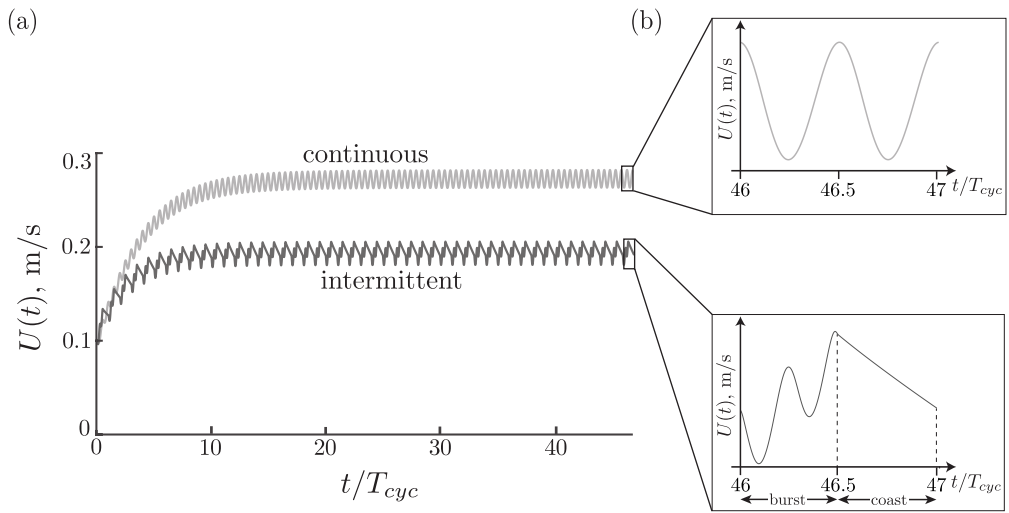}
\caption{(a) Swimming speed as a function of the nondimensional time normalized by the period of a cycle for the continuous or intermittent swimmer.  (b) Swimming speed as a function of nondimensional time over a single cycle of motion with the continuous swimmer in the top inlay and the intermittent swimmer in the bottom inlay.}
\label{velocity}
\end{figure}

Figure \ref{velocity}a presents the swimming speed as a function of time for a representative continuous and intermittent swimmer.  Both swimmers have the same Lighthill number of $Li = 0.04$, and are operating with the same frequency and amplitude of $f = 1$ Hz and $A^* = 0.7$, respectively.  The swimmers only differ in their duty cycle with the continuous swimmer using $DC = 1$ and the intermittent swimmer using $DC = 0.5$.  The swimmers start with an initial velocity of $U = 0.1$ m/s and accelerate up to a steady-state condition after 10 -- 20 cycles.  Predictably, the intermittent swimmer has a lower mean speed at steady-state than the continuous swimmer.  Figure \ref{velocity}b shows the velocity as a function of time only over a single cycle for both swimmers.  The continuous swimmer has the typical pattern observed in self-propelled studies \cite[]{Moored2017} where there are two surging oscillations about the mean speed in a given cycle.  This correlates with the two vortex shedding events and the two thrust peaks observed in oscillatory pitching motions.  In contrast, the intermittent swimmer produces a different swimming pattern.  During the bursting phase of swimming there are two surging oscillations superposed onto an acceleration of the intermittent swimmer.  During the coasting phase the swimmer is decelerating under the prescribed $U^2$ drag law with no oscillations in speed since the hydrofoil propulsor is static.

\begin{figure}
\centering
\includegraphics[width=0.9\textwidth]{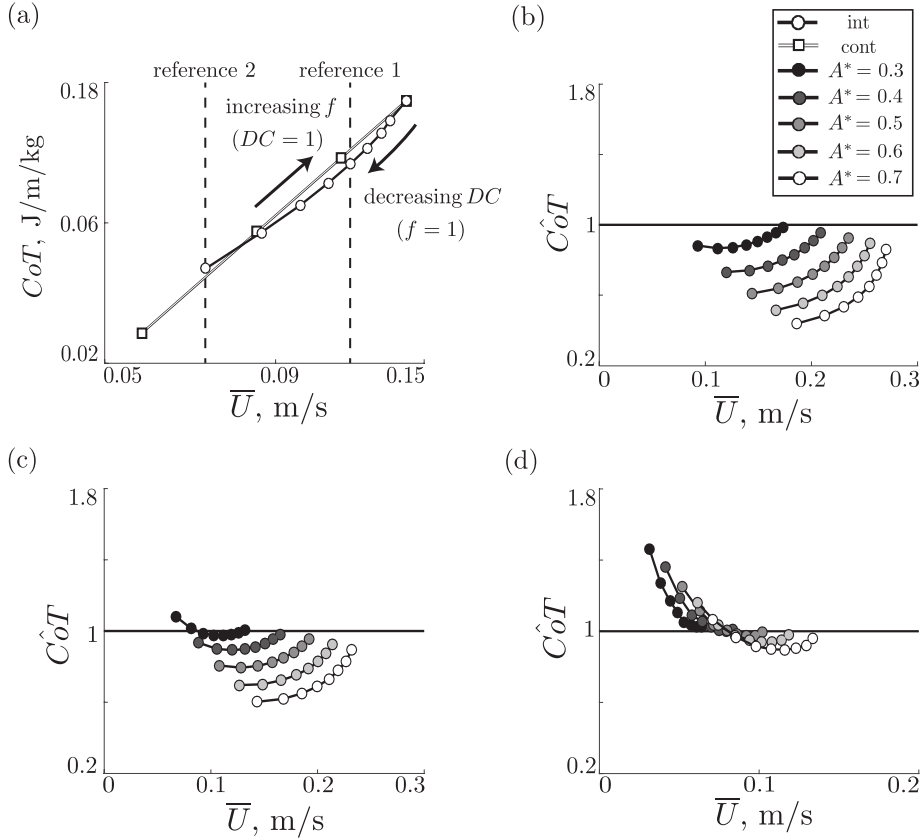}
\caption{(a) Cost of transport as a function of the mean speed for a continuous and intermittent swimmer.  The swimming parameters are $A^*=0.7$ and $Li=0.4$ for both swimmers.  Normalized cost of transport as a function of the mean speed for three Lighthill numbers: (b) $Li=0.04$, (c) $Li=0.08$, and (d)  $Li=0.4$.}
\label{performance}
\end{figure}

Figure \ref{performance}a presents the $CoT$ as a function of the mean speed for a continuous and intermittent swimmer on a log-log scale.  For both swimmers their amplitude of motion and Lighthill number are $A^*=0.7$ and $Li = 0.4$, respectively.  The continuous swimmer has a fixed duty cycle of $DC = 1$ and increases its oscillation frequency from $f =0.2$ -- $1$ Hz.  As the frequency of motion increases the mean speed increases linearly and the $CoT$ increases quadratically, which is made evident by the slope of the continuous swimmer line being equal to two on a log-log scale. 

In contrast, the intermittent swimmer has a fixed frequency of $f = 1$ Hz and decreases its duty cycle from $DC = 1$ -- $0.2$.  At $DC = 1$ and $f = 1$ Hz, the intermittent and continuous swimmers are equivalent.  When the duty cycle is decreased, the mean speed of the intermittent swimmer drops and consequently the $CoT$ decreases.  It can be observed that there is a range of $DC$ where by using an intermittent gait swimmers can lower their $CoT$ while maintaining their swimming speed as compared to a continuous swimming gait (reference line 1), albeit with an increased oscillation frequency during the burst period.  Additionally, there is a range of $DC$ where by using an intermittent gait costs more energy than continuous swimming at the same mean speed (reference line 2).  Importantly, the energetically favorable regime of intermittent swimming is observed with \textit{potential flow} simulations that do not compute the skin friction from the boundary layer, but instead rely on a drag law with a \textit{constant} drag coefficient.  This means that the observed benefit in the computations cannot be due to the viscous Bone-Lighthill boundary layer thinning hypothesis \cite[]{Lighthill1971}, which relies on an \textit{increase} in the skin friction drag during oscillation to make intermittent swimming energetically favorable over continuous swimming.  Instead, since a beneficial regime is still observed with potential flow computations then it must be due to an inviscid mechanism.  This leads to a host of questions such as, how much of a benefit is observed?  In what range of parameters does this benefit exist?  What is the inviscid mechanism?  How do the forces and energetics of intermittent swimming scale? What role does viscosity and the Bone-Lighthill mechanism play?  The first four of these questions are examined in the current study while the last question is the subject of future work.

Figures \ref{performance}b-d show the normalized cost of transport as a function of the mean swimming speed for three different Lighthill numbers, that is, $Li = 0.04,\, 0.08,$ and $0.4$.  If $\hat{CoT} > 1$ then a continuous swimming gait is energetically beneficial.  If $\hat{CoT} = 1$ then both gaits are energetically equivalent.  Finally, if $\hat{CoT} < 1$ then an intermittent gait is energetically beneficial.  Regardless of the $Li$, choosing intermittent swimming is energetically more favorable as the amplitude of pitching increases. Additionally, the available energetic benefit from intermittent swimming becomes greater for decreasing $Li$ of swimmers. In fact, at $Li > 0.4$ there is essentially a negligible energetic savings with only minute savings at very high duty cycles that can be garnered by switching to an intermittent gait.  However, the $Li$ limit can be recast into a limit on the Reynolds number of swimming.  Based on the drag coefficient of the laminar boundary layer on a flat plate $C_D = 0.664\, Re_l^{-1/2}$ and the area ratio of $S_{wp} = 8$, then $Li = 5.312\, Re^{-1/2}$ and intermittent swimming would produce no observable energetic benefits for $Re < 175$.  At least this is an estimate for the inviscid benefit of intermittent swimming, but a Bone-Lighthill viscous benefit is not necessarily restricted in the same manner. 

The lower bound on $Re$ where an intermittent benefit can be observed is reflected in biological studies. Larval northern anchovy uses continuous swimming whereas adult northern anchovy use an intermittent swimming mode \cite[]{Weihs1980,Muller2000}. As larval northern anchovy reach $5$ mm in length their typical Reynolds number during swimming is $Re \approx 100$ and coincidently the fish begin to swim intermittently. 

For the lowest $Li$ and highest amplitude examined in this study, the $CoT$ reduction is as high as 60\%. In other words, a swimmer can choose to swim intermittently rather than continuously and save as much as 60\% of its energy per unit mass to travel a unit distance at the same speed. Importantly, unlike previous studies, this conclusion is obtained by \textit{fixing} the drag coefficient for both bursting and coasting periods of swimming, meaning that the observed benefit is not due to the Bone-Lighthill boundary layer thinning hypothesis, but instead there is an inviscid mechanism that leads to the energy savings.

\subsection{Wake Dynamics}
\begin{figure}
\centering
\includegraphics[width=0.95\textwidth]{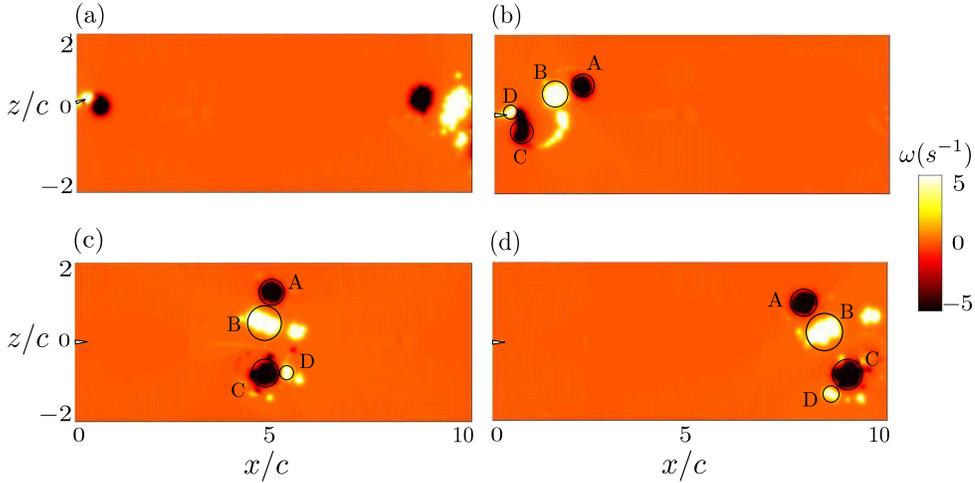}
\caption{Evolution of the vorticity field of an intermittently pitching hydrofoil with $DC=0.2$, $A^*=0.7$ and $Li=0.08$. The evolution of the vortex wake is shown at times, (a) $t/T_{cyc} = 1/20$, (b) $t/T_{cyc} = 1/5$, (c) $t/T_{cyc} = 3/5$, and (d) $t/T_{cyc} = 1$.}
\label{generic_flow_field}
\end{figure}

\begin{figure}
\centering
\includegraphics[width=0.95\textwidth]{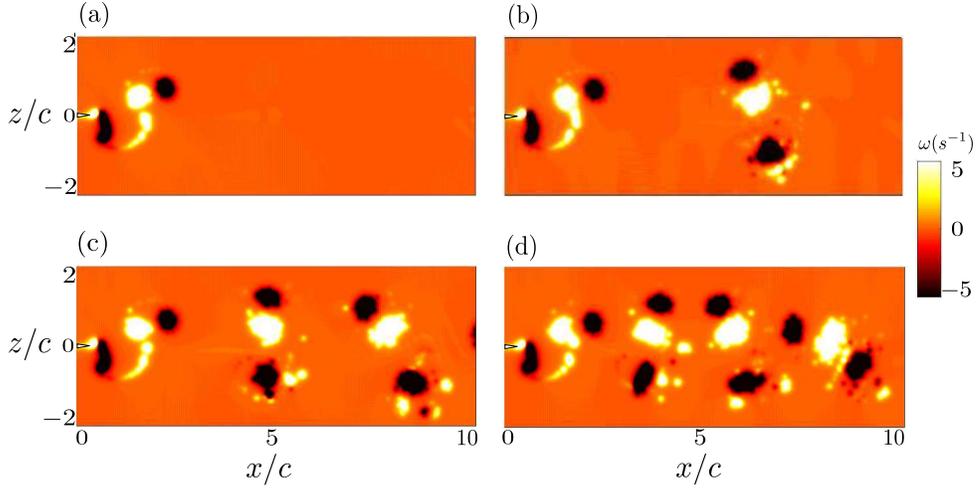}
\caption{Vorticity field comparison of hydrofoils pitching with $A^*=0.7$, $Li=0.08$. Each frame shows a different duty cycle with (a) $DC=0.2$, (b) $DC=0.4$, (c) $DC=0.6$, and (d) $DC=0.8$.}
\label{DC_similarity}
\end{figure}


 Figure \ref{generic_flow_field} shows the evolution of the vorticity field throughout a cycle for an intermittent swimmer with $DC=0.2$. The distribution of positive (anti-clockwise) and negative (clockwise) vorticity is represented with white and black, respectively. There is no near wake at the beginning of the burst period since previous wake structures have advected multiple chord lengths away (Figure \ref{generic_flow_field}a). Over each bursting period, four vortices are shed; one starting vortex (vortex A), two vortices are shed as the hydrofoil changes direction (vortices B and C) and one weak stopping vortex (D) (Figure \ref{generic_flow_field}b). Vortices B and C are stronger and form a pair as they travel downstream (Figures \ref{generic_flow_field}c and \ref{generic_flow_field}d).  Due to mutual induction, vortices B and C advect downstream faster than vortices A and D. On the other hand, vortex A induces some vorticity from vortex B to be left behind in its downstream propagation (Figure \ref{generic_flow_field}d). 

\begin{figure}
\centering
\includegraphics[width=1.0\textwidth]{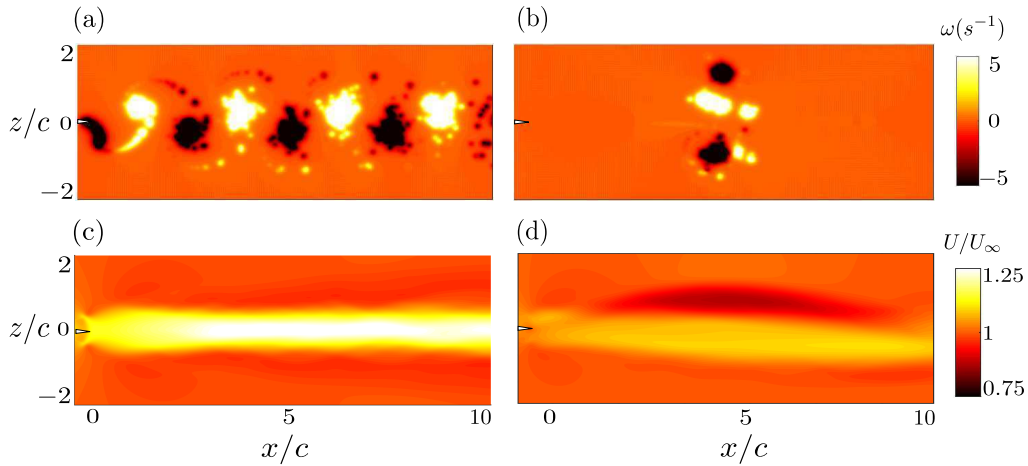}
\caption{The vorticity field of hydrofoils pitching with $A^*=0.7$ and $Li=0.08$ at (a) $DC = 1$ and (b) $DC=0.2$. Time averaged velocity fields of (c) the continuous and (d) the intermittent gait at a mean speed of $\overline{U} = 0.14$ m/s.}
\label{average_jet}
\end{figure}

 Figure \ref{DC_similarity} shows the self-similarity of the vortex groups shed during the bursting period when the duty cycle is altered. For all duty cycles examined four vortices are shed per bursting period in the manner described above.  As observed in Figure  \ref{DC_similarity} the evolution of the vortex groups are effectively unaltered by the duty cycle. However, the spacing of the vortex groups grows inversely proportional to the duty cycle.  The self-similarity of vortex groups produced by an intermittent swimmer breaks down for the limiting case of continuous swimming.  Figure \ref{average_jet}a shows a typical reverse \vK vortex street where two vortices are shed per pitching cycle produced by a continuous swimmer as observed previously \cite[]{Freymuth1988,Koochesfahani1989}.

The induced velocities among the three strong vortices shed during a bursting period of an intermittent gait (Figure \ref{average_jet}b) leads to an asymmetry in the time-averaged velocity field. Figure \ref{average_jet}c and \ref{average_jet}d show the comparison of the time-averaged $x$-velocity field of a continuously pitching hydrofoil ($f = 0.58$ Hz and $DC = 1$) with its intermittently pitching counterpart swimming at the same mean speed of $\overline{U}=0.14$ m/s. For the continuous swimmer, the symmetrically shed vortices create a reverse von K\'{a}rm\'{a}n street, which time-averages to a single-core jet aligned in the streamwise direction. For the intermittent swimmer, in the near wake ($x/c \leq 1$), the vortices are shed in a similar manner to a classic reverse \vK street creating a time-averaged momentum jet aligned in the streamwise direction. In the far wake ($x/c \geq 1$), the three strong vortices evolve into two pairs with one pair inducing a velocity in the downstream direction while the other induces a velocity upstream.  In the time-averaged sense this creates a momentum surplus and deficit branch that are parallel to each other for $2 \leq x/c \leq 8$.  The momentum jet is observed to also have an asymmetry suggesting that there is finite time-averaged lateral force produced by the swimmer.


\subsection{Scaling Laws of Intermittent Swimming} \label{sec:scale}
In this section the scaling of the forces and energetics of intermittent swimming with the swimming parameters will be considered.  Of prime importance are the scaling of the mean swimming speed and the cost of transport, which are inherently linked to the scaling of the mean thrust production and mean power consumption of a swimmer \cite[]{Moored2017}.  Here, we will show that a simple scaling of the thrust and power with the duty cycle can be used to transform the intermittent swimming problem into a continuous swimming problem.  Then previously derived scaling relations for continuous swimming will be shown to seamlessly apply to scaling the transformed thrust and power.  

\begin{figure}
\centering
\includegraphics[width=0.95\textwidth]{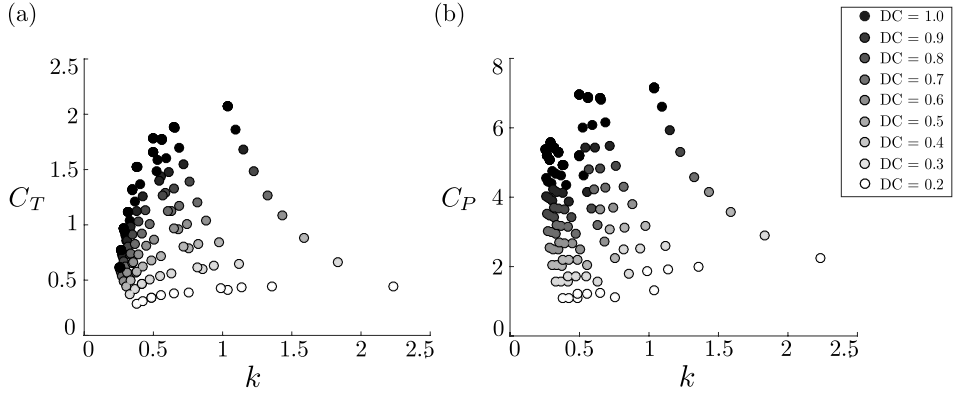}
\caption{Mean (a) thrust coefficient and (b) power coefficient as a function of reduced frequency.}
\label{Ct_Cp_unscaled}
\end{figure}

Figure \ref{Ct_Cp_unscaled}a shows the thrust coefficient for swimmers as a function of the reduced frequency for various duty cycles ranging from $0.2 \leq DC \leq 1$.  The mean thrust used in the definition of the thrust coefficient is averaged over an entire cycle, that is, the combined bursting and coasting periods.  For a given duty cycle, the thrust is observed to increase with an increase in the reduced frequency.  Similarly, for a fixed reduced frequency the thrust increases with an increase in the duty cycle.  Figure \ref{Ct_Cp_unscaled}b presents the power coefficient as a function of the reduced frequency.  Like the thrust coefficient, the power is seen to increase with increasing reduced frequency and duty cycle.  The thrust and power both show a large range of values with the variation of the parameters $f,\, Li,\, DC,$ and $A^*$.  

\begin{figure}
\centering
\includegraphics[width=0.95\textwidth]{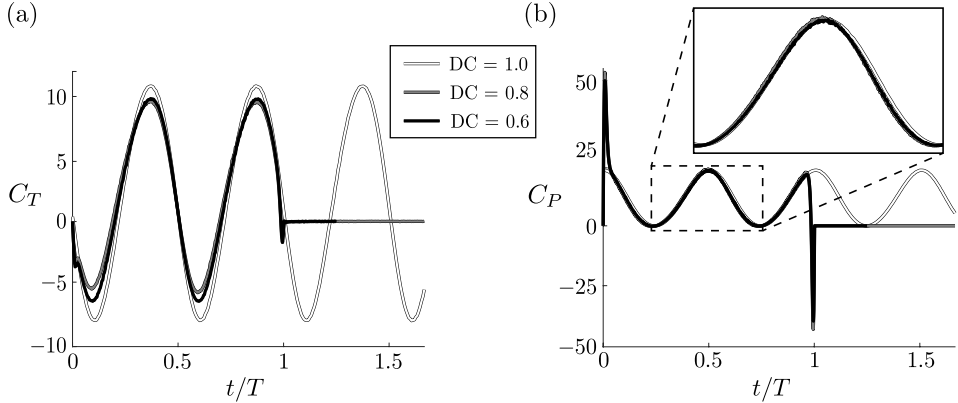}
\caption{Instantaneous (a) thrust coefficient  and (b) power coefficient for three duty cycles of $DC=0.6$, $0.8$, and $1$.  The frequency, amplitude and Lighthill number are $f = 1$ Hz, $A^* = 0.7$, and $Li = 0.04$, respectively, for all of the duty cycles.}
\label{thrust_over_cycle}
\end{figure}

To gather more insight into how the thrust and power scale with the kinematic parameters, the instantaneous thrust and power are presented in Figure \ref{thrust_over_cycle}.  The thrust for the continuous swimmer shows two characteristic peaks that are associated with the peak angular accelerations that occur during a cycle and the subsequent shedding of two vortices per oscillation cycle.  Similarly, the intermittent swimmers with $DC = 0.6$ and $0.8$ show two peaks in thrust that are associated with the shedding of the two strongest vortices that form during an intermittent swimming cycle.  In contrast, there are two small troughs in the instantaneous thrust at the beginning and end of the burst period that are associated with shedding of the starting and stopping vortices that occur from the high accelerations at the beginning and end of the burst period.  As the duty cycle is lowered from $DC = 1$ the peaks and troughs of the instantaneous thrust are attenuated with the troughs showing large reductions in their peak drag.  By considering only the bursting period, it is evident that the mean thrust actually rises as the duty cycle decreases.  However, when considering the mean thrust over the entire cycle, this small rise in thrust during the bursting period is overcome by the effectively zero thrust throughout the coasting phase.  It is now clear that the near zero thrust during the coasting phase coincides with the reduction in the mean thrust (Figure \ref{Ct_Cp_unscaled}a) and the subsequent reduction in the mean swimming speed (Figure \ref{performance}a) as the duty cycle is lowered.  Since the thrust is zero over the coasting phase a better comparison of the mean thrust among different duty cycles of swimming would average the thrust only over the bursting period, $C_{T,\text{burst}}$.  The mean thrust over the bursting period and the mean thrust over the entire cycle are simply connected as,
\begin{align} \label{CTscale}
\quad \quad C_{T,\text{burst}} = \frac{C_T}{DC}.
\end{align} 

\noindent The major effect of the thrust reduction with decreasing duty cycle can then be accounted for by scaling the mean thrust over the entire cycle with the duty cycle as in (\ref{CTscale}).  

In Figure \ref{thrust_over_cycle}b the continuous swimmer is seen to have two peaks in power over an oscillation cycle that occur near the peak angular velocity of the pitching hydrofoil.  In contrast, the intermittent swimmers have a prominent peak and trough in the power consumption at the beginning and end of the bursting period in accordance with the high accelerations at those times, however, they essentially cancel and provide a negligible contribution to the mean power.  As the duty cycle decreases the instantaneous power is seen to be amplified only slightly over the bursting period and consequently the mean power over the bursting phase is observed to increase minutely for this case.  The instantaneous power also shows zero power consumption during the coasting phase since the pitching hydrofoil has no angular velocity during that phase.  Like the thrust, this leads to a reduction in the mean power over the entire cycle, which can be accounted for by averaging the power over the bursting period or equivalently by scaling the power with the duty cycle,
\begin{align} \label{CPscale}
\quad \quad C_{P,\text{burst}} = \frac{C_P}{DC}.
\end{align} 

Now, the thrust and power coefficients are scaled in accordance with (\ref{CTscale}) and (\ref{CPscale}) and presented in Figure 11a and 11b.  The thrust data has now collapsed to a line while the power data is more compact, but there is still stratification in the data.  What is particularly striking is that the data now resembles self-propelled data from only continuous swimmers {\cite[]{Moored2017}.  This shows that by applying the simple scalings in (\ref{CTscale}) and (\ref{CPscale}), the intermittent swimming problem has been \textit{transformed} into a continuous swimming problem.  It is now hypothesized that previously derived scaling relations for self-propelled continuous swimmers can account for the physics of the transformed problem.
\begin{figure}
\centering
\includegraphics[width=0.95\textwidth]{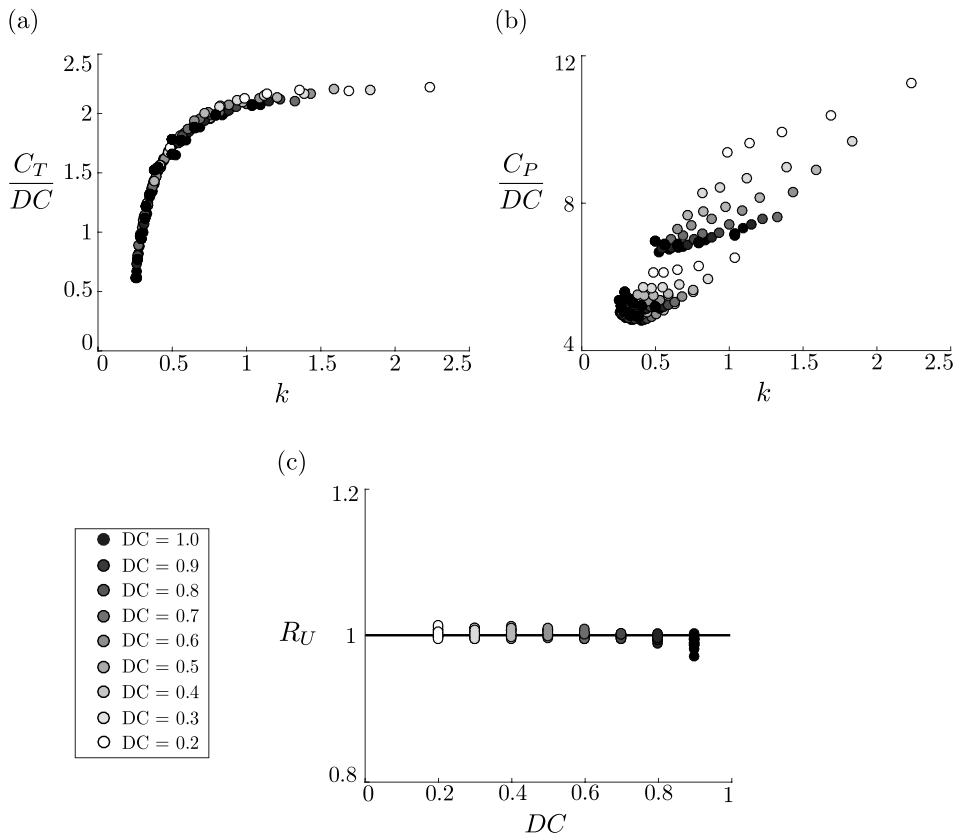}
\caption{(a) Mean scaled thrust coefficient as a function of the reduced frequency. (b) Mean scaled power coefficient as a function of the reduced frequency. (c) Ratio of the predicted mean speed from the scaling relation to the mean speed from the simulations as a function of $DC$.}
\label{Ct_and_Cp}
\end{figure}

Moored \& Quinn (\citeyear[]{Moored2017}) examined the inviscid scaling laws of a self-propelled pitching hydrofoil combined with a virtual body (same model as the current study).  The nonlinear numerical results were compared with the exact scaling relations from Garrick's linear theory \cite[]{Garrick1936}.  Garrick's theory extends Theodorsen's theory \cite[]{Theodorsen1949} by determining the power consumption of a heaving and pitching hydrofoil and by accounting for the singularity in the vorticity distribution at the leading-edge in order to calculate the thrust.  Both approaches are linear theories and assume that the hydrofoil is infinitesimally thin, that there are only small amplitudes of motion and that the wake is planar and non-deforming.  Surprisingly, even with these assumptions the scaling of the thrust coefficient from Garrick's linear theory was found to capture the scaling behavior of the mean thrust from nonlinear numerical results \cite[]{Moored2017}.  The thrust scaling relation for continuous swimming proposed by Moored \& Quinn ({\citeyear[]{Moored2017}) is then a Garrick-inspired relation with
\begin{align} \label{Ct_cont}
C_{T}^{\text{cont}}(k) &= c_1 \, - c_2 \, w(k) \\
 \mbox{where } w(k) &= \frac{3F}{2} + \frac{F}{\pi^2 k^2} - \frac{G}{2\pi k} - \left( F^2 + G^2 \right)\left( \frac{1}{\pi^2 k^2} + \frac{9}{4} \right) \nonumber
\end{align}

\noindent Here $F$ and $G$ are the real and imaginary parts of Theodorsen's lift deficiency function \cite[]{Theodorsen1949} and $c_1$ and $c_2$ are coefficients that are to be determined from the data.  The wake function $w(k)$ is specified for a pitching axis at the leading edge.  Now, the more general intermittent swimming problem is transformed into a continuous swimming problem with $C_T/DC = C_T^{\text{cont}}$ from (\ref{CTscale}) giving the following generalized scaling for the thrust coefficient,
\begin{align} \label{Ct_scale_int}
\frac{C_{T}(k)}{DC} &= c_1 \, - c_2 \, w(k)
\end{align}

\noindent From this relation it is evident that by plotting $C_T/DC$ on a vertical axis and $k$ on a horizontal axis the data should collapse to a line that is only a function of $k$.  This is precisely what is observed in Figure \ref{Ct_and_Cp}a, which confirms that the scaling relation (\ref{Ct_scale_int}) captures the proper physics.

The physical meaning of the first and second terms on the right hand side of (\ref{Ct_scale_int}) can be directly interpreted from Garrick's theory where they represent the added-mass and circulatory streamwise forces, respectively.  In fact, for pitching about the leading edge, the circulatory term is always drag inducing for all $k$ and the added-mass term is always thrust producing.  Consequently, as the reduced frequency of a swimmer is increased, they will increase the ratio of their thrust producing added-mass forces relative to their drag inducing circulatory forces.  

Through a simple algebraic extension the thrust scaling relation can be used to calculate the self-propelled mean swimming speed of an intermittent swimmer.  By recognizing that during steady-state swimming the mean thrust is balanced by the mean drag then the thrust relation can be combined with the drag law (\ref{eq:draglaw}) to determine the mean swimming speed,
\begin{align} \label{U_scale}
\overline{U}_{\text{scale}}=fA \sqrt{2\,DC\frac{C_T^{\text{cont}}(k)}{Li}}.
\end{align}

\noindent To make this scaling relation predictive only two simulations would need to be run to determine the coefficients in (\ref{Ct_cont}).  Here, the coefficients are determined to be $c_1 = 0.536$ and $c_2 = 2.3$ by minimizing the squared residuals with the entire set of simulation data.  Now, the scaling relation prediction can be directly assessed against the data gathered from the numerical simulations by forming a ratio of the mean swimming speeds as, $R_U=\overline{U}_{\text{scale}}/\overline{U}_{\text{sim}}$.  If this ratio is equal to one then the scaling relation provides an exact prediction of the mean speed from the nonlinear simulations.  Figure \ref{Ct_and_Cp}c presents the ratio of swimming speeds as a function of the duty cycle.  It is evident that the scaling relation provides an excellent prediction of the mean swimming speed for an intermittent swimmer where the mean speed scaling prediction is within 3\% of its full-scale value from the simulations.  Here the drag law used in the simulations was prescribed \textit{a priori}, however, the key to predicting the swimming speed is to have the proper thrust scaling relation, which is not prescribed \textit{a priori} in the simulations.  This result confirms the algebraic extension from thrust to swimming speed and that the thrust scaling is accurate for a range of reduced frequencies.  

In contrast to the thrust scaling relation, Moored and Quinn (\citeyear[]{Moored2017}) determined that Garrick's linear theory did \textit{not} accurately capture the scaling trends in the power coefficient for their nonlinear simulation data.  They argued that linear theory needed to be augmented by two nonlinear corrections that are not accounted for in linear theory due to its assumptions.  Their proposed power scaling relation for continuous swimming is the following,
\begin{align} \label{CP_scale}
C_P^{\text{cont}}(k,St) &= c_3 + c_4\, \frac{St^2}{k} \left(\frac{k^*}{k^*+1}\right) + c_5\, St^2k^*\\
 &\mbox{where }  k^* \equiv \frac{k}{1 + 4\, St^2} \nonumber
\end{align}

\noindent Here, the first term on the right hand side of (\ref{CP_scale}) represents the added mass power directly from linear theory while the second and third terms represent the nonlinear corrections.  The second term occurs since the large amplitude pitching motions exhibit a finite streamwise component of velocity that through the vortex force produces lift and an additional power term.  Importantly, due to the presence of a nonlinearly separating shear layer at the trailing edge there is a phase shift in the bound circulation development \cite[]{Wang2013,Liu2014} that allows some terms to be nonorthogonal when determining the mean power.  Due to these physics the second term is denoted as the large amplitude separating shear layer term.  The third term is known as the vortex proximity term and it is due to the vortex force that occurs when the trailing-edge vortex induces a streamwise component of velocity over the hydrofoil.  
\begin{figure}
\centering
\includegraphics[width=0.95\textwidth]{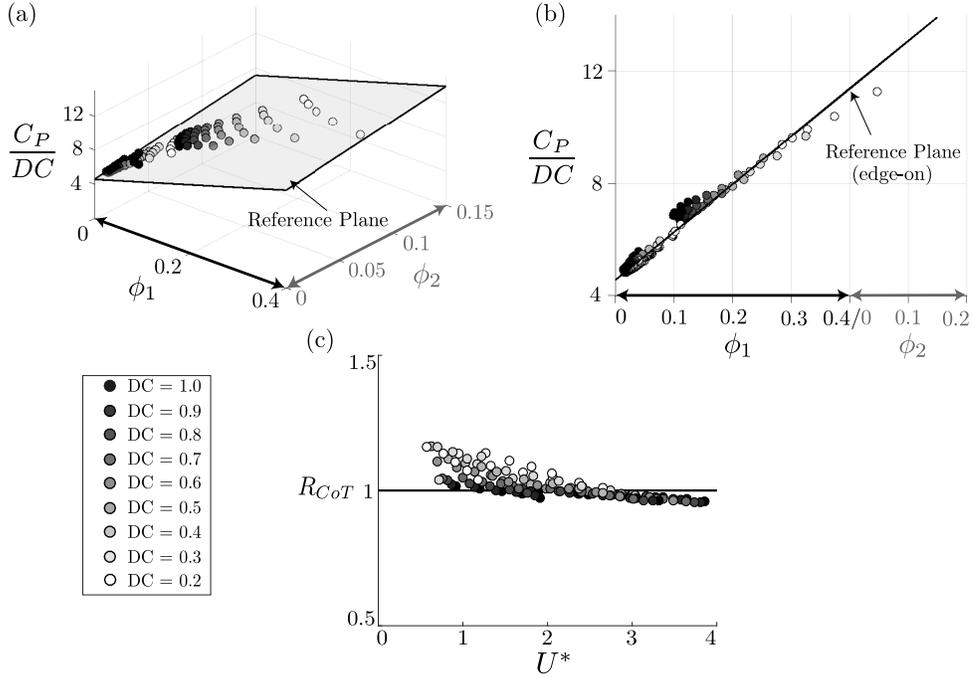}
\caption{The scaled mean power coefficient as a function of $\phi_1$ and $\phi_2$ from (a) a perspective view and (b) an ``edge-on" view of the reference plane.  (c) Ratio of the cost of transport predicted from the scaling laws to the actual computational data as a function of the duty cycle.}
\label{Scaling_vs_simulation}
\end{figure}

Now, the power scaling relation for continuous swimming (\ref{CP_scale}) can be generalized for intermittent motions by applying the transform (\ref{CPscale}), which gives
\begin{align} \label{CP_scale_int}
\frac{C_P(k,St)}{DC} &= c_3 + c_4\, \frac{St^2}{k} \left(\frac{k^*}{k^*+1}\right) + c_5\, St^2k^*.
\end{align}

\noindent Here the coefficients $c_3$, $c_4$, and $c_5$ are to be determined from the numerical data.  By examining (\ref{CP_scale_int}) it becomes evident that if $C_P/DC$ is plotted as a function of $\phi_1 = (St^2/k) \left(k^*/(k^* + 1)\right)$ and $\phi_2 = St^2 k^*$ in a three-dimensional graph, then the numerical data should collapse to a plane.  Figure \ref{Scaling_vs_simulation}a shows a perspective view of the three-dimensional data and Figure \ref{Scaling_vs_simulation}b shows an ``edge-on" view of the data with a reference plane.  It can be seen that there is an excellent collapse of the scaled power coefficient data to a plane.  This confirms that the continuous swimming scaling relations account for the physics of intermittent swimming once the intermittent transformation is applied.  

The power coefficient scaling relation can be extended through simple algebra to calculate a scaling relation for the cost of transport giving
\begin{align}\label{CoT_scale}
CoT_{scale} = \frac{\rho S_p f^2 A^2}{m} DC\, C_P^{\text{cont}}(k,St). 
\end{align}

To make this scaling relation predictive, only three simulations would need to be run to determine the three coefficients, $c_3$, $c_4$, and $c_5$.  Here the coefficients are determined to be $c_3 = 4.551$, $c_4 = 17.08$, and $c_5 = 16.86$ by minimizing the squared residuals for the entire data set. Subsequently, the ratio of the predicted cost of transport compared to the cost of transport from the simulations can be formed as $R_{CoT}=CoT_{\text{scale}}/CoT_{\text{sim}}$.  Figure \ref{Scaling_vs_simulation}c shows the ratio of the cost of transports as a function of the nondimensional speed where $R_{CoT} = 1$ represents a perfect scaling prediction of the actual simulation data.  The scaling prediction is observed to predict the actual cost of transport to within 18\% of its full-scale value.  This confirms the algebraic extension of the power coefficient scaling to the cost of transport and that the power scaling relation is accurate for a range of $k$ and $St$.

\subsection{Inviscid Energy Saving Mechanism}

Section \ref{sec:perf} presented the result that by using intermittent motions instead of continuous motions a swimmer can save energy when swimming at the same mean speed in an \textit{inviscid} environment.  The energy reduction was observed to: (1) increase with increasing amplitudes of motion, (2) increase with decreasing $Li$, and (3) only occur over a limited range of $DC$ outside of which it would cost additional energy to swim intermittently rather than continuously.  Now, the scaled data from Section \ref{sec:scale} can be examined to elucidate the inviscid mechanism behind the energetic benefit of intermittent swimming and consequently to understand these three phenomena that determine the limits of the energy savings.
\begin{figure}
\centering
\includegraphics[width=0.95\textwidth]{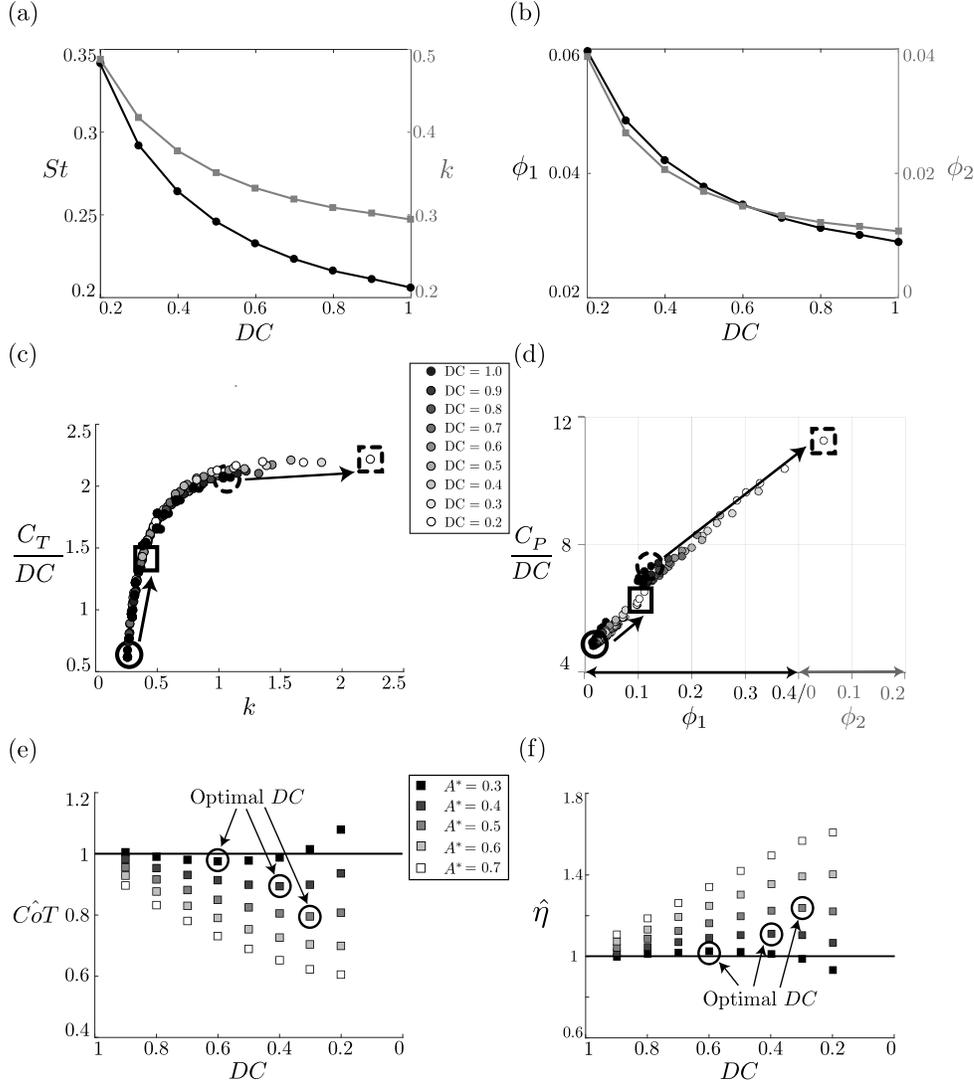}
\caption{(a) Strouhal number and reduced frequency as a function of the duty cycle for $A^* = 0.7$ and $Li = 0.08$. (b) Large-amplitude separating shear layer term and the vortex proximity term as a function of the duty cycle for $A^* = 0.7$ and $Li = 0.08$.  (c) Scaled thrust as a function of reduced frequency.  (d) Scale power as a function of $\phi_1$ and $\phi_2$.  (e) Normalized cost of transport as a function of duty cycle.  (f) Normalized efficiency as a function of duty cycle.  The black circles indicate the optimal duty cycle for each $A^*$.}
\label{mechanism}
\end{figure}

When a swimmer switches from continuous to intermittent swimming and reduces its duty cycle it will consequently swim slower (Figure \ref{performance}).  If the frequency of motion remains constant then the reduced swimming speed will result in a rise in $k$ and $St$.  As expected, Figure \ref{mechanism}a presents a monotonic increase in the reduced frequency and Strouhal number as the $DC$ decreases for $A^* = 0.7$ and $Li = 0.08$.   At the same time, the large-amplitude separating shear layer term, $\phi_1 = (St^2/k) \left(k^*/(k^* + 1)\right)$, and the vortex proximity term, $\phi_2 = St^2 k^*$, also show a monotonic increase with decreasing duty cycle (Figure \ref{mechanism}b).  By examining the scaled thrust and power coefficients (Figure \ref{mechanism}c and \ref{mechanism}d) it is clear that since $k$, $\phi_1$, and $\phi_2$ all increase monotonically with decreasing $DC$ the thrust and power coefficients will also increase with decreasing $DC$.  However, it is observed that the potential increase in the scaled thrust coefficient becomes negligible at sufficiently high $k$ while there is no such trend in the scaled power coefficient.  

To examine this more closely, a high amplitude swimmer with $A^* = 0.7$, $Li = 0.04$, and $DC = 1$ is highlighted on both the scaled thrust and power data by solid black circles Figure \ref{mechanism}c and \ref{mechanism}d.  The swimmer then reduces its duty cycle to $DC = 0.2$ and its new scaled thrust and power conditions are highlighted on the graphs by solid black squares.  It is observed that the thrust coefficient rises by 113\% ($C_T/DC = 0.6734$ to $1.434$) while the power coefficient rises by only 7.5\% ($C_P/DC = 5.05$ to $5.46$).  Since the propulsive efficiency (\ref{eff_cot}) can be restated as $\eta = C_T/C_P$ then in this case, where the thrust increase outweighs the power increase, there is a subsequent rise in the propulsive efficiency of 97.7\% ($\eta=13.3\%$ to $26.3\%$).  There is a second case of a low amplitude swimmer with $A^* = 0.3$, $Li = 0.4$, and $DC = 1$ that is highlighted on both the scaled thrust and power data by dashed black circles.  The swimmer then lowers its duty cycle to $DC = 0.2$ and its new scaled thrust and power conditions are highlighted on the graphs by dashed black squares.  It is observed that when the $DC$ decreases the thrust coefficient only rises by 6.7\% ($C_T/DC = 2.071$ to $2.219$) while the power coefficient rises by 53.6\% ($C_P/DC = 7.33$ to $11.26$).  Then it is evident that the power increase outweighs the thrust increase and the propulsive efficiency drops by 30.4\% ($\eta=28.3\%$ to $19.7\%$).  

To connect these findings with the energy savings results presented in Section \ref{sec:perf} the propulsive efficiency and cost of transport are presented as the normalized efficiency and cost of transport (see (\ref{Norm_cot_eff})) for the same $A^*$ and $Li$ (Figure \ref{mechanism}e and \ref{mechanism}f).  If $\hat{\eta} > 1$ intermittent swimming is more efficient than the continuous swimming and \textit{vice versa}.  Similarly, if $\hat{CoT} < 1$ intermittent swimming uses less energy than the continuous swimming and \textit{vice versa}.  Furthermore, the efficiency and cost of transport can be directly linked \textit{at the same mean speed} by recognizing that during steady-state swimming the mean thrust and drag must balance, which leads to $\overline{T}=1/2\,C_d \rho S_p U^2$ and
\begin{align}
CoT= \frac{\overline{P}}{m\overline{U}} = \frac{\overline{P} \overline{T}}{m\overline{U}\overline{T}} = \frac{\frac{1}{2}C_d\rho U^2}{m\, \eta} = \frac{\text{const.}}{\eta}.
\end{align}

\noindent This relation shows that at the same mean speed, $CoT \propto 1/\eta$ and Figure \ref{mechanism}f is just the inverse of Figure \ref{mechanism}e.  It can be observed that there is an optimal duty cycle marked by the black circles in Figure \ref{mechanism}e and \ref{mechanism}f that minimizes the normalized cost of transport and maximizes the normalized efficiency simultaneously. Additionally, the optimal duty cycle decreases with increasing amplitude of motion.

Now, it is evident that energy savings occur when a continuous swimmer has a sufficiently low reduced frequency in the range of $0.25 \leq k \leq 1$ such that by reducing their $DC$ will result in a thrust increase that will outweigh their power increase.  Consequently, their efficiency will increase and their cost of transport will decrease.  If a swimmer has a high reduced frequency of $k > 1$ then lowering their $DC$ will result in a power increase that outweighs their thrust increase, which will lead to lower efficiency and higher cost of transport.  

Physically, as $DC$ is reduced, $k$ increases and the swimmer is increasing the ratio of added mass thrust producing forces to the circulatory drag inducing forces to improve their thrust.  At sufficiently high $k$ the thrust coefficient is dominated by the added mass forces and there are no further gains that can occur by reducing the $DC$.  In contrast, the power coefficient shows an ever increasing trend with decreasing $DC$.  Therefore, the inviscid mechanism behind the energy savings observed in intermittent swimming can be described as a \textit{Garrick mechanism}, which relies on the trade-off between the added mass thrust producing forces and the circulatory drag inducing forces and it is independent of the Bone-Lighthill viscous mechanism.  Future work will examine the details of intermittent swimming in a viscous flow to help disentangle the contributions to the energy savings from the inviscid Garrick mechanism and the viscous Bone-Lighthill mechanism.

By understanding the inviscid Garrick mechanism behind the energy savings of intermittent swimming, the three observed phenomena stated at the beginning of this section can now be understood.  First, using high amplitude motions results in low reduced frequencies during self-propelled swimming and \textit{vice versa}.  This mean that the greatest energy savings can occur for high amplitude swimmers.  Second, swimmers with low Lighthill numbers will also have low reduced frequencies during self-propelled swimming and subsequently they have the potential for the greatest energy savings.  Finally, the energy savings of intermittent swimming switches to an additional energetic cost over continuous swimming when the force production of the swimmer is dominated by added mass forces, which leaves little to gain in thrust and large additional costs in power as the swimmer's duty cycle is reduced.
 

\section{Conclusion}
The performance and wake structures of a self-propelled swimmer modeled as a combination of a virtual body and an intermittently pitching hydrofoil are examined by using an inviscid boundary element method.  In contrast with previous studies, a drag law is prescribed for the virtual body that has a \textit{constant} drag coefficient, that is, it does not depend upon whether a swimmer is bursting or coasting.  Under these conditions it is discovered that swimming intermittently can save as much as 60\% of the energy it takes to swim continuously at the same speed even in an inviscid environment.  This finding shows that in addition to the viscous Bone-Lighthill boundary layer thinning mechanism proposed by \cite{Lighthill1971} there is another, potentially more important, mechanism behind the energy savings of intermittent swimming.  

The inviscid mechanism is described as a Garrick mechanism where the added-mass thrust-producing forces can be increased relative to the circulatory drag-inducing forces by increasing the reduced frequency during self-propelled swimming.  A key discovery is that a self-propelled intermittent swimmer can directly increase its reduced frequency by reducing its duty cycle.  Additionally, the energy savings during intermittent swimming are shown to (1) increase for increasing amplitude of motion, (2) increase for decreasing $Li$, and (3) switch to an additional energetic cost above continuous swimming at sufficiently low $DC$.  The first two observations occur since both high amplitude and low $Li$ continuous swimmers have low reduced frequencies during self-propelled swimming.  This allows for the greatest potential gains in the thrust coefficient as the $DC$ is decreased before the thrust coefficient saturates at sufficiently high $k>1$.  At the same time the power coefficient increases, but is outweighed by the thrust coefficient increase, resulting in an increase in efficiency.  The third observation occurs when a swimmer has reached a sufficiently high $k$ such that by decreasing the duty cycle further the thrust coefficient effectively does not change while the power coefficient still increases, leading to a decrease in efficiency.  

In contrast to continuous swimming, it is observed that during intermittent swimming four vortices are shed per pitching cycle regardless of the duty cycle.  More precisely, there are two vortices shed when the hydrofoil changes its pitching direction as well as a starting and a stopping vortex shed at the beginning and the end of the bursting period.  In the time average, the mean streamwise velocity field shows an asymmetric jet structure with both momentum surplus and deficit branches.

It is further discovered that if the thrust and power coefficients are averaged over only the bursting period then the intermittent swimming problem can be transformed into a continuous swimming problem.  This is equivalent to normalizing the thrust and power coefficients averaged over the entire cycle by the $DC$.  Previous continuous swimming scaling laws were then generalized to include intermittent swimming by incorporating the scaling transformation. The thrust and power scaling relations show excellent collapse of the data to a line and a plane, respectively, confirming that the underlying physics are accurately accounted for in the relations.  Finally, the thrust and power scaling relations are algebraically extended to calculate the mean swimming speed and the cost of transport.  By tuning a few coefficients with a handful of simulations these relations can become predictive.  In the current study, the mean speed and cost of transport are predicted to within 3\% and 18\% of their full-scale values by using these relations.

\section*{Acknowledgments}
This work was funded by the Office of Naval Research under Program Director Dr B. Brizzolara, MURI Grant Number N00014-14-1-0533.


\bibliography{Intermittent_Swimming_Paper.bib}
\bibliographystyle{jfm}

\end{document}